# Orientation of the northern gate of the Goseck Neolithic rondel

Marianna Ridderstad

University of Helsinki, Observatory, P. O. Box 14, FI-00014 University of Helsinki, Finland; email: ridderst@kruuna.helsinki.fi

**Abstract**

The two southernmost gates of the Goseck rondel, built by the *Stichbandkeramik* culture around 4800 BCE, were oriented to the winter solstice sunrise and sunset. The northern gate of the rondel deviates a few degrees from the Meridian line. It is suggested that this deviation from the cardinal direction is due a stellar orientation towards Edasich, which was the pole star at the time, and is thus related to the concept of the world pillar and the pole star as the top of it.

Over 200 Neolithic rondels have been found in the area extending from Northern Germany to Hungary. Many of them show passage orientations symmetrically towards either the cardinal or the intermediate (northeast-southwest and northwest-southeast) directions. Astronomical orientations have been found for three rondels in Sachsen-Anhalt (Schwarz 2004), and there are also strong indications that the many rondels built by the partly contemporaneous Lengyel culture in the Carpathian basin are similarly related to astronomical observations (Pásztor and Barna 2008). The gates of the rondels, for which astronomical orientations have been determined, are typically oriented towards either the sunrises and sunsets on the key solar dates of the year (Becker 1996; Karlovský and Pavúk 2002; Kovárník et al. 2006; Schwarz 2004; Pásztor and Barna 2008).

The Goseck rondel consists of a circular ditch (about 71 m across) and two wooden palisade rings (56 m and 49 m across) (Bertemes et al. 2004). It is believed that the Goseck rondel was used as a solar observatory and a place of worship by the *Stichbandkeramik* (hereinafter STK) culture (Bertemes et al. 2004; Schlosser 2004), which inhabited the area between 4900 BCE and 4500 BCE (Stadler 1995).

The astronomical function of the rondel has been deduced from the orientations of the southeast (SE) and southwest (SW) gates of the Goseck rondel, which are towards the midwinter sunrise and sunset, respectively, in 4800 BCE (Schlosser 2004). The Goseck rondel also has a third, northern passage. It is not oriented straight to the north, but with an error of a few degrees (see Figure 1). This kind of small deviation from the Meridian line is also seen in many other rondels (see, e.g., Kovárník et al. 2006:Figure 3).

In Neolithic structures, the direction of the true north was probably often estimated simply by dividing in two the angle between the rising and setting points of the sun (Burl 1995:21). This method often leads to errors of many degrees, when the horizon is not even, since the azimuth of the sun changes when it rises higher. However, the northern gate of Goseck is not on the line going through the middle point obtained using the two southernmost gates, which were the midwinter solar rising and setting points at the time when the structure was built.

The cultures that built the first rondels, the *Linearbandkeramik*, STK and Lengyel cultures, were some of the first farmer cultures in Northern and Central Europe. Their relation to the sun, the life-giver of the farming society, must have been affected by the significance of the sun to the agricultural activities. This explains the solar orientations of the rondels. On the other hand, as common as the cardinal orientations for prehistoric enclosures like rondels and henges are, the actual meaning of them is not clear. A strong candidate for the significance of the Meridian orientations is the concept of *axis mundi*, the sky-supporting pillar, and the top of it, the pole star.

In 4800 BCE, the closest star to the celestial pole brighter than 4 mag was *Edasich* (ι Draconis; 3.28 mag), which was located about 5 degrees from the north celestial pole. At midwinter, before sunrise, when the sun was at -14 degrees below the horizon line, Edasich was at (azimuth 8 degrees, altitude 51 degrees).[1] Since Edasich was so close to the celestial pole, it was, when observed from the center of the structure, visible above the northern gate for several hours before the sunrise on the day of the winter solstice, when the structure had its two other orientations.[2] Thus, the northern gate could be directed to Edasich, the pole star.

---

[1] The first/last appearance of stars brighter than 3 mag that are 60 degrees or farther from the sun in azimuth, and higher than about 10 degrees above the horizon, has been estimated to occur when the altitude of the sun is at –10 degrees below the horizon line. For a 4 mag star at the same altitude and farther than about 120 degrees from the sun in azimuth, the first/last appearance occurs when the sun is about –14 deg below the horizon line (see Purrington 1988).

The "pole star" or the "nail star", being attached to the head of the sky-supporting "world pillar", holds a strong position in the ancient beliefs of many North-Eurasian people (see Setälä 1932; Eliade 1987; Hultkrantz 1996 and refs. therein). According to Setälä (1932:539-41, 547, 560), the "nail" was usually described as being made of iron, but in the Finnish and Estonian mythology, it is also described as "golden", which probably is the older epithet of these two. Not only the nail, but also the whole pillar was described as being made of iron, gold, or copper by many North-Eurasian peoples.

There are historical accounts that both the North-European Germanic tribes and the Samí had, in their houses and shrines, sacred wooden columns, which had nails attached to the top of them (Setälä 1932:537-48). In Germany, Scandinavia and Lapland, the wooden columns in the temples and outdoor shrines represented the world pillar that supports the sky. The nails of the Vikings were called "the nails of gods" or "the world nails", whereas the Samí nails were more clearly related to the pole star, which was, and still is, called *boahenavlle*, "the nail of the bottom" or "the nail of the north". The sky was seen as a giant concave "lid", which was attached to the sky pillar with the nail at its bottom. This worldview is also seen in the ancient Finnish name of the sky, *kirjokansi*, "the embroidered lid". The lid was supported by *sammas* or *sampo*, the metal world pillar, which had a golden nail atop of it.

Parpola (2004) suggested that the Finnish and Estonian word *sampo* for the world pillar, around which the sky revolves, and which has the "nail star" atop of it, is of Proto-Indo-European origin from 3000-2000 BCE. At this time, the pole star was Thuban ($\alpha$ Draconis), which had moved closer than Edasich to the north celestial pole around 3900 BCE. The North-Eurasian concept of the world pillar as a revolving entity could thus have originated with the Proto-Indo-Europeans using the early chariot wheel, which was invented in 4000-3000 BCE. The North-Eurasian distribution of the "pole star" –myth also points towards a common origin.

However, the myth of a world pillar without the metal "nail" atop of it is probably much older than 4000 BCE. As it is also encountered among some Native American tribes (Hultkrantz 1996), it is probably of ancient Paleolithic origin.

The religion of many North Eurasian and North American tribes, which had the world pillar concept as an important part of their beliefs, was essentially shamanistic (Hultkrantz 1996). It has been argued, based on representations in cave art, that the Upper Paleolithic religion in Europe was shamanistic (Clottes and Lewis-Williams 1998), and that also the early Neolithic religious practices had shamanistic elements (Lewis-Williams and Pearce 2005).

If Edasich as the pole star, pointing the way to the north celestial pole, was important in Northern Europe in about 5000 BCE, it may have been related to the Paleolithic shamanistic concept of the world pillar. In 4500-4000 BCE, when the early use of metals first began in Central Europe (Höppner et al. 2005), the myth may have included the pole star seen as the metal top of the pillar, which later merged with the idea

---

[2] The center of the Goseck rondel has been taken to be the point where the observations of winter solstice sunrise and sunset have been made (see Figure 1).

of the early wheel axle. Finally, the metal pillar-top specified into a nail, a common object.

Along with the North-Eurasian concept of the world pillar came the belief that the edge of the sky could move up and down in the horizon (see Hultkrantz 1996). This movement allowed people, both alive and dead, to pass into the otherworld. The shaman was the one person alive who was able to assist the dead on their way to the otherworld and also visit the other realms to gain knowledge useful for the community (see, e.g., Hoppál 2005). In Northern Europe, the late autumn and winter have been seen as the dark time related to the dead and the netherworld. The orientations of the Goseck rondel to both the winter solstice and the pole star could be related to these kinds of shamanistic beliefs of its builders.

It is known that in the first cemeteries of Northern European hierarchical societies in the Mesolithic, the burials of shamans differed from those of all others (O'Shea and Zvelebil 1984). In this context, the relatively few burials in STK rondels (Meyer and Raetzel-Fabian 2006) can be explained as resulting from the custom of placing only the graves of the shamans, the religious elite, inside the rondels.

In conclusion, it seems that the orientations of the three passages of the Goseck rondel tell about the mixed system of beliefs after the Neolithic revolution, which merged the old Paleolithic shamanistic beliefs, including the concept of the sky-supporting pillar, with the new ones holding the sun in central position.

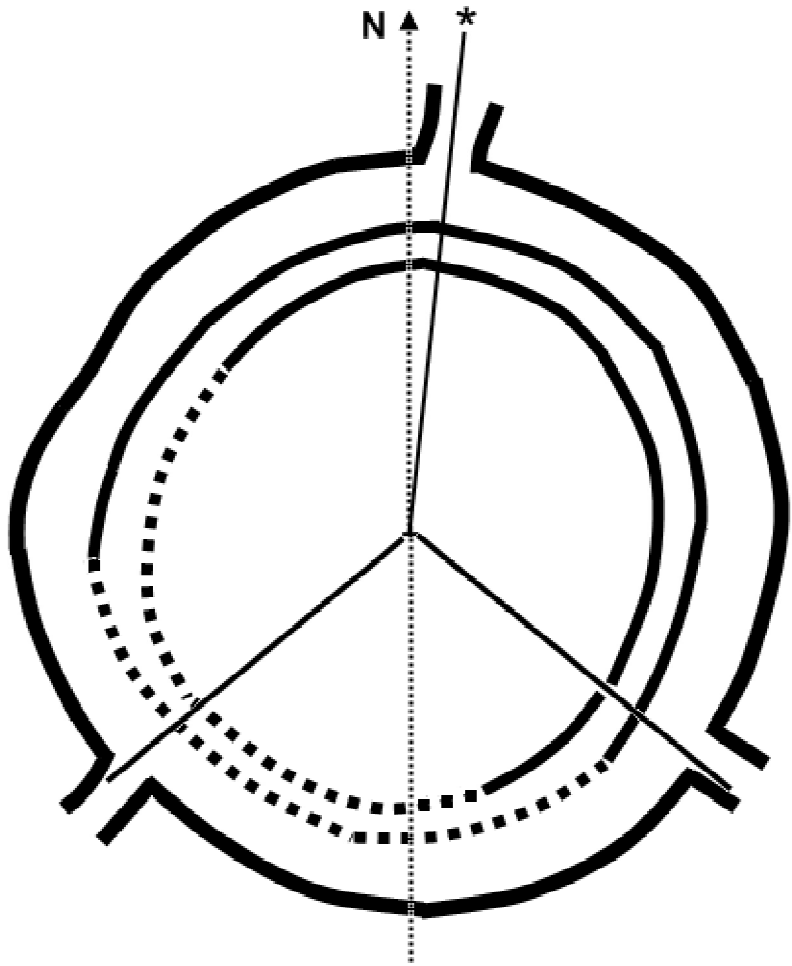

**Figure 1.** The Goseck prehistoric rondel consists of two wooden palisade rings inside a circular ditch. The SE and SW gates of the structure are oriented towards the sunrise and sunset, respectively, at the winter solstice in 4800 BCE, whereas the orientation of the northern gate deviates 3-9 degrees from true north. Edasich, the pole star around 4800 BCE, was seen through the northern gate before the sunrise in the winter solstice. (The map of the rondel redrawn by the present author from the archaeoastronomical analysis photograph on the Internet site of the Goseck multimedial archaeology project by Institut für Prähistorische Archäologie, Landesamt für Archäologie Sachsen-Anhalt, and Multimedia Authoring Center for Teaching in Anthropology (University for California Berkeley): http://www.praehist.uni-halle.de/goseck/index2.htm)